\documentclass[12pt]{iopart}

\usepackage{graphicx}
\usepackage{iopams}

\newcommand{\Mpl}{M_{\rm{Pl}}}
\newcommand{\kl}[1]{{\textstyle #1}}
\newcommand{\BmL}{$B$$-$$L$}
\newcommand{\Hinf}{H_{\rm Inf}}

\begin{document}

\title{\BmL-symmetric Baryogenesis
with Leptonic Quintessence}

\author{Mathias Garny}

\address{Max-Planck-Institut f\"ur Kernphysik, P.O. Box 103980,
D 69029 Heidelberg, Germany}
\ead{mathias.garny@mpi-hd.mpg.de}

\begin{abstract}

We discuss a toy model where baryogenesis and cosmic acceleration
are driven by a leptonic quintessence field coupled to
the standard model sector via a massive mediating scalar field. It does not require the
introduction of \BmL-violating interactions below the inflationary scale.
Instead, a \BmL-asymmetry is stored in the quintessence field,
which compensates for the corresponding observed baryon asymmetry.

\end{abstract}


\section{Introduction}

Scalar fields with time-dependent vacuum expectation value are commonly invoked in cosmology,
above all to
describe the inflationary phase~\cite{Guth:1980zm} of the early universe, and they are also considered in dynamical
dark energy models, called quintessence models~\cite{Wetterich:1987fm,Ratra:1987rm}, aiming to explain the apparent present 
acceleration~\cite{Riess:1998cb,Perlmutter:1998np} of our cosmos. Furthermore, rolling fields are the basis of 
a number of baryogenesis models~\cite{Affleck:1984fy,Dine:2003ax} and also play an important role in the context of a possible time-variation
of fundamental constants over cosmological time-scales~\cite{Uzan:2002vq}. Due to the similarity of the underlying concepts,
it is an interesting question whether some of these issues could be related. 
This has been studied for example for the early- and late time acceleration,
called quintessential inflation~\cite{Peebles:1998qn}, or for the
combination of spontaneous lepto- and baryogenesis with quintessence~\cite{Li:2001st,Yamaguchi:2002vw}
and quintessential inflation~\cite{DeFelice:2002ir}. 

Complex scalar fields have also been discussed
as candidates for dynamical dark energy~\cite{Boyle:2001du,Gu:2001tr},
which offers the possibility that the field carries a $U(1)$-charge,
and thus could 
itself store a
baryon or lepton density~\cite{Bauer:2005vz}.
This approach can very well be accommodated  within the so-called
``baryosymmetric 
baryogenesis''~\cite{Dodelson:1989ii,Dolgov:1991fr} scenario, where
one attempts to explain the overabundance of matter over antimatter
without introducing new baryon- (B) or lepton (L) number violating interactions,
nevertheless starting with no initial asymmetry. This requires the introduction
of an invisible sector, in which an asymmetry is hidden that exactly compensates the
one observed in the baryon (and lepton) sector, thereby circumventing one of the Sakharov conditions~\cite{Sakharov:1967dj}. 
Here we will review a possible realization where the anomaly-free combination
\BmL\ is conserved below the inflationary scale, and the invisible sector which compensates for the
\BmL-asymmetry of the standard model (SM) baryons and leptons is leptonic dark energy following Ref.~\cite{Bauer:2005vz}.
For other realizations involving dark matter
or neutrinos see e.g. Refs.~\cite{Dodelson:1989ii,Dick:1999je}.

%
\section{Toy Model}
%

In this section we address the question how \BmL-asymmetries
in the dark energy sector, realized by a complex quintessence field charged under \BmL, and in the SM sector can be
created by a dynamical evolution within an underlying
\BmL-symmetric theory.
For this, it is necessary to consider a suitable interaction
between both sectors. Direct couplings
between the quintessence field and SM fields are commonly investigated for example in the context of
time-varying coupling constants and/or~-masses~\cite{Uzan:2002vq} or violations
of the equivalence principle~\cite{Ratra:1987rm}, which leads to strong constraints
in the case of a coupling e.g. to photons or nucleons~\cite{Carroll:1998zi,Ratra:1987rm,Garny:2006wc}.
Here, we discuss a toy model where we assume that direct interactions between the
quintessence field $\phi$ and the SM are sufficiently suppressed, allowing however an indirect interaction
mediated by a ``mediating field'' $\chi$ which couples to $\phi$ and the SM. In the late universe,
the $\chi$-interactions should freeze out which means that
the massive scalar $\chi$ is hidden today and also that
the transfer of asymmetry between the quintessence 
and the SM sector freezes out. Thus, once an asymmetry
has been created in each sector in the early universe, it will not be washed out
later on. In the specific setup considered here we take the quintessence field
to carry lepton number $-2$, so that it carries a \BmL-density given by
\begin{equation}\label{BmLdensity}
n_\phi = -2|\phi|^2\dot\theta_\phi \qquad (\textrm{with}\ \phi\equiv |\phi| e^{i\theta_\phi}),
\end{equation}
and analogously for the mediating field $\chi$ which carries the same lepton number.
The effective toy-model Lagrangian for $\phi$ and $\chi$ we consider is
\begin{eqnarray*} \mathcal{L} &=& \kl{\frac{1}{2}}(\partial_{\mu}\phi)^{*}(\partial^{\mu}\phi)-V(|\phi|)
 + \kl{\frac{1}{2}}(\partial_{\mu}\chi)^{*}(\partial^{\mu}\chi)-\kl{\frac{1}{2}}\mu_{\chi}^{2}|\chi|^{2}\\
&& -  \kl{\frac{1}{2}}\lambda_{1}|\phi|^{2}|\chi|^{2}-\kl{\frac{1}{4}}\lambda_{2}(\phi^{2}\chi^{*2}+\mbox{h.c.})
 + \mathcal{L_{SM}}(\chi,\dots),
\end{eqnarray*}
with dimensionless coupling constants $\lambda_1>0$ and $\lambda_2<\lambda_1$ responsible for the coupling
between the quintessence and the mediating field.
This Lagrangian has a global $U(1)$-symmetry under a common phase rotation of $\phi$ and $\chi$
which corresponds to a \BmL-symmetric theory. The coupling of the mediating field
to the SM encoded in the last contribution should also respect this symmetry. This is compatible
e.g. with a Yukawa-like coupling of the form $\mathcal{L}_{SM}\ni -g\chi\overline{\nu_R^c}\nu_R +\mbox{h.c.}$
to right-handed neutrinos, see Ref. \cite{Bauer:2005vz} for a more detailed discussion.
For the quintessence potential we assume an exponential potential of the 
form~\cite{Barreiro:1999zs,Ferreira:1997au,Wetterich:1987fm,Ratra:1987rm}
$V(|\phi|)=V_{0}\left(e^{-\xi_{1}|\phi|/\Mpl}+ke^{-\xi_{2}|\phi|/\Mpl}\right)$
which leads to tracking of the dominant background component
and a crossover towards an accelerating attractor at the present epoch for $\xi_1\gg 2 \gg \xi_2$ 
and a suitable choice of $k$~\cite{Barreiro:1999zs}. For the dynamics in the early universe
one can safely neglect the second term. 
Since the vacuum expectation value (VEV) of $\phi$ increases and typically $|\phi|\gtrsim\Mpl$ today, the effective mass
$m_\chi^2\approx\mu_\chi^2+\lambda_1|\phi|^2$ of the mediating field gets huge
and the field indeed decouples the quintessence and the SM sectors in the late
universe. However, before the electroweak phase transition the dynamics of $\phi$ and $\chi$
can lead to a creation of the baryon asymmetry.

%
\section{Creation of a \BmL-asymmetry}
%

\begin{figure}
\includegraphics[clip,scale=0.6]{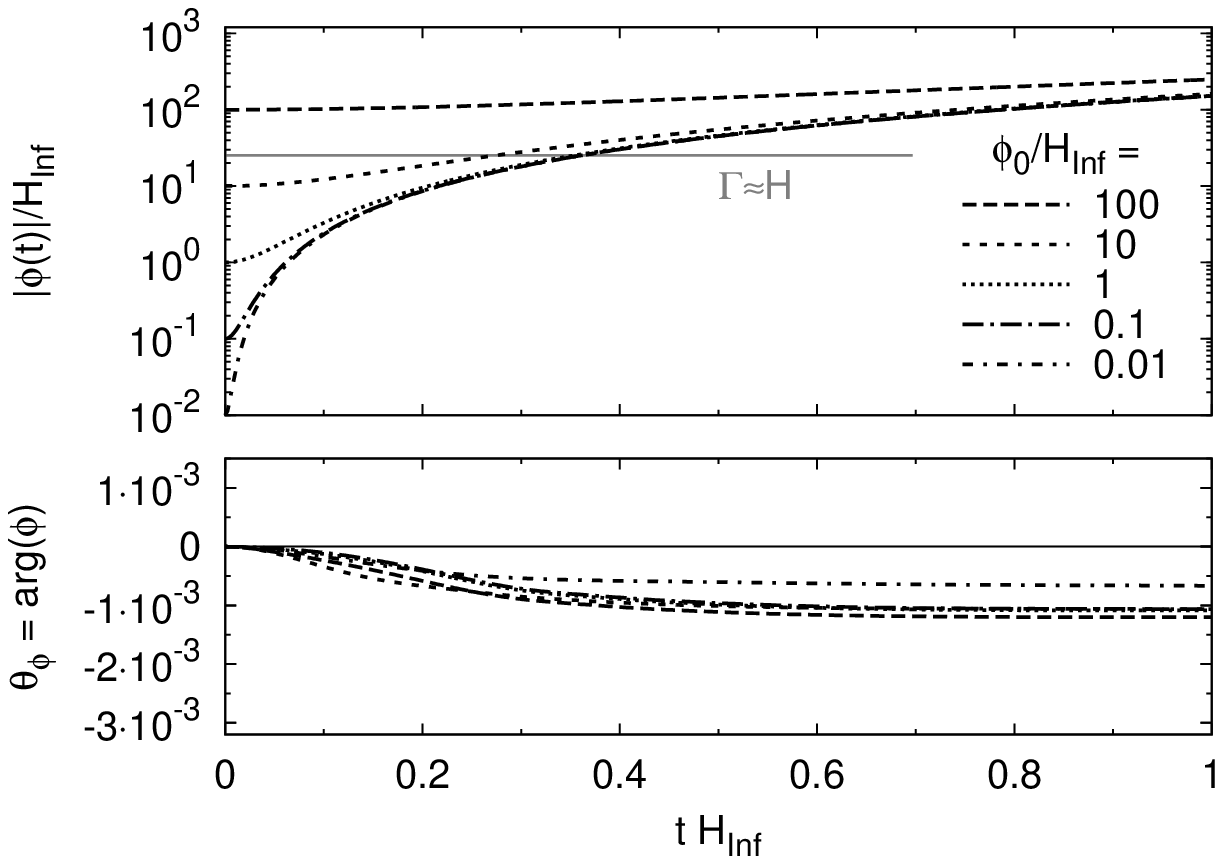}
\includegraphics[clip,scale=0.6]{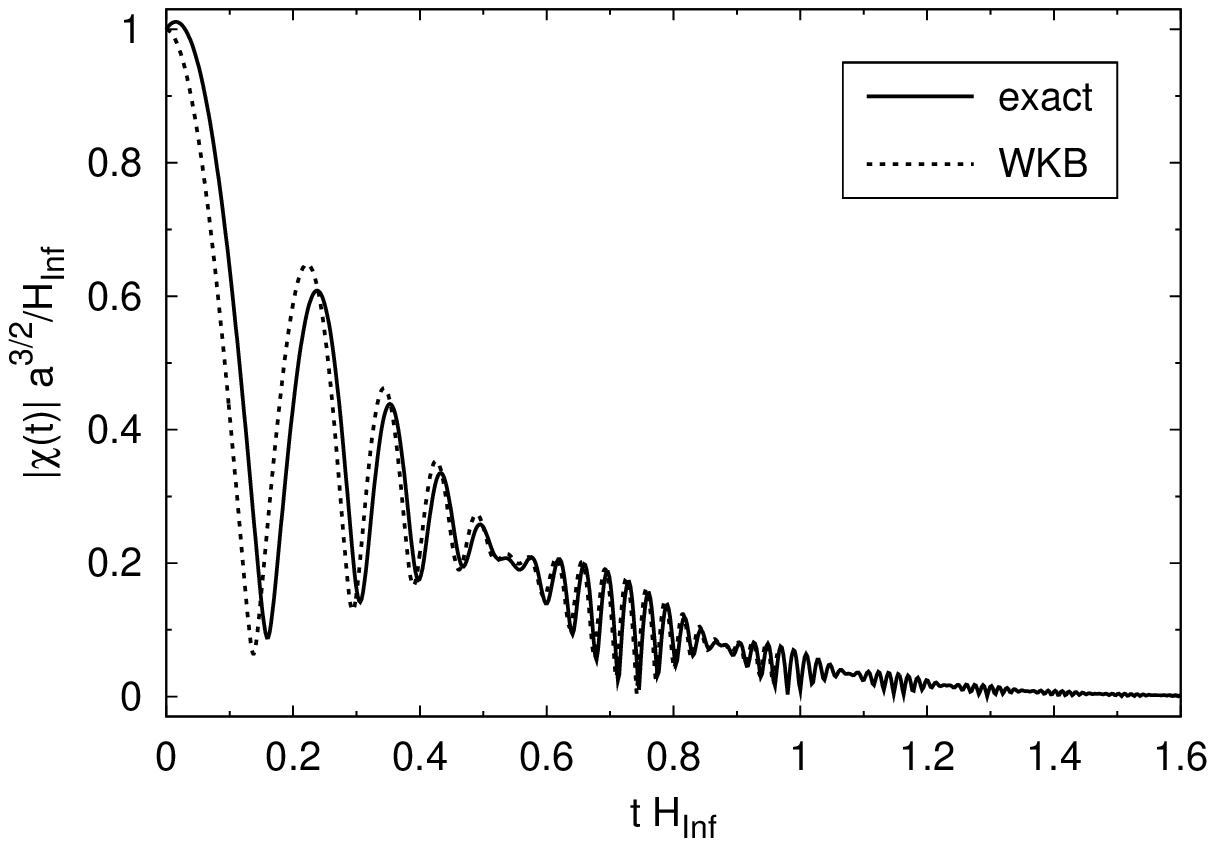}
\caption{\textbf{Left:} Numerical solution for the absolute value of the quintessence VEV $|\phi|$ (upper) and
its complex phase (lower) for various initial conditions $\phi_0$ and the choice 
$\lambda_1=1, \lambda_2=0.1, V_0/\rho_0=10^{-5},\xi_1 = 7, \chi_0=H_{\rm Inf}=10^{12}{\rm GeV},\alpha_0=\frac{\pi}{4}, g=1$
of parameters. \textbf{Right:} Numerical and approximate WKB solution for the absolute value of the mediating field VEV $|\chi|$
for the same parameter values despite $\phi_0=H_{\rm Inf}$.}\label{PhiAndChi}
\end{figure}

\begin{figure}
\begin{center}
\includegraphics[clip,scale=0.5]{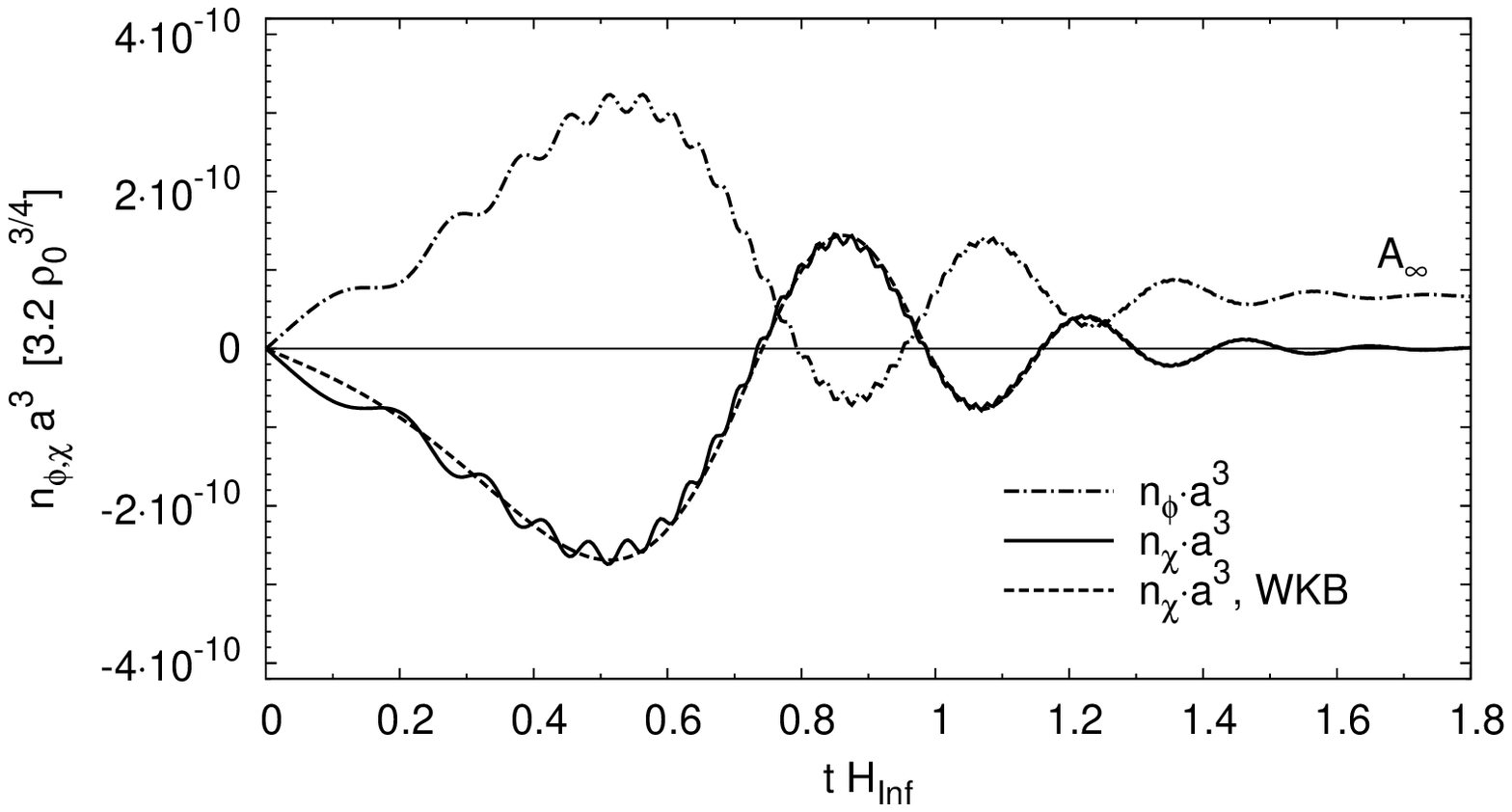} \hspace{-0.2cm}
\includegraphics[clip,scale=0.53]{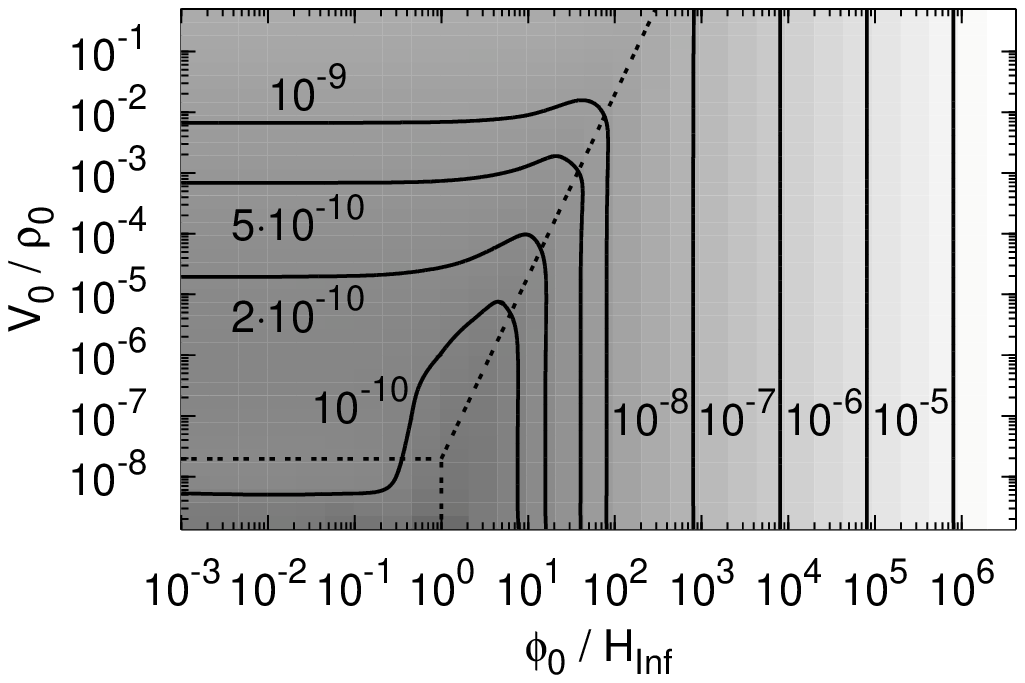}
\end{center}
\caption{\textbf{Left:} Time-evolution of the comoving asymmetry of the quintessence (dot-dashed) and the mediating (solid) fields for the
same parameters as in Fig.~\ref{PhiAndChi} despite $g=0.5$. After an
initial phase of oscillations, the $\chi$-field decays and the asymmetry stored in the quintessence
field goes to a constant asymptotic value $A_\infty$ which is of equal amount but opposite sign as the
asymmetry created in the SM. The analytic WKB approximation for $n_\chi$ is also shown (dashed).
\textbf{Right:} Contour plot of the created asymmetry $\kappa\propto A_\infty$. $V_0/\rho_0$ corresponds
to the fraction of quintessence energy density after inflation and $\phi_0$ is the initial quintessence VEV. 
The other parameters are chosen as in Fig.~\ref{PhiAndChi}. 
The dashed lines divide the regions where the analytic approximations from eq. (\ref{kappaWKB}) are valid. }\label{Asym}
\end{figure}

To study the evolution of the scalar fields $\phi$ and $\chi$ in the early
universe, we assume that it is described by a flat FRW metric
after the end of inflation with a Hubble parameter $H=\Hinf$ and with
VEVs $\phi=\phi_0$ and $\chi=\chi_0 e^{-i\alpha_0}$ inside our Hubble patch
which are displaced by a relative angle $\alpha_0$ in the complex plane.
These initial conditions correspond to dynamical $CP$ violation if $\sin (2\alpha_0)\not= 0$, 
which is necessary for the formation of an asymmetry~\cite{Balaji:2004xy,Dolgov:1991fr}.
Under these conditions, the fields start rotating in the complex plane
and thus develop a \BmL-density, see eq. (\ref{BmLdensity}). This asymmetry
is then partially transfered to the SM by the \BmL-conserving decay of the $\chi$-field into SM particles,
leading to a decay term  for the $\chi$-field
in the equations of motion~\cite{Bauer:2005vz}
\begin{eqnarray*}
\ddot{\phi}+3H\dot{\phi} & = & -2\frac{\partial V}{\partial \phi^{*}}-\lambda_{1}|\chi|^{2}\phi-\lambda_{2}\phi^{*}\chi^{2},\label{eomphi}\\
\ddot{\chi}+3H\dot{\chi}+3\Gamma_{\chi\rightarrow{\rm SM}}\dot{\chi} & = & -\mu_\chi^2\chi-\lambda_{1}|\phi|^{2}\chi-\lambda_{2}\chi^{*}\phi^{2},
\end{eqnarray*}
where $\Gamma_{\chi\rightarrow{\rm SM}}=\frac{g^2}{8\pi}m_\chi$ is the decay rate and $g^2$ stands for the squared sum
of the Yukawa couplings corresponding to the relevant decay channels. Provided that the
quintessence behaviour is dominated by the exponential and not by the mixing terms
(which is roughly the case if $|V'(\phi_0)|\gg \chi_0^2\phi_0, \chi_0^3$),
it will roll to larger field values with only small changes in the radial direction, whereas
the $\chi$-field oscillates and decays once $\Gamma_{\chi\rightarrow{\rm SM}}\gtrsim H$ (see Fig. \ref{PhiAndChi}).
Due to the \BmL-symmetry, the total \BmL-density is conserved,
and thus the asymmetries stored in the different components always
add up to the initial value which we assume to be zero after inflation, i.e.
\begin{equation}
n_\phi + n_\chi + n_{\rm SM} \equiv 0.
\end{equation} 
After the decay of the $\chi$-field,
the comoving asymmetry freezes (see left part of Fig. \ref{Asym}) since there is no more exchange between the 
quintessence and the SM sectors\footnote{We set $t\equiv 0$, $a\equiv 1$ at the end of inflation}~\cite{Bauer:2005vz},
\begin{equation}\label{Ainfty}
n_{\rm SM} a^3 \rightarrow -n_\phi a^3 \rightarrow {\rm const} = 
\int_0^\infty\!\!\!\! \textrm{d}t\,a^3 \Gamma_{\chi\rightarrow{\rm SM}}\cdot n_\chi \equiv A_\infty,
\end{equation}
and thus the \BmL-asymmetry in the SM is exactly compensated by the \BmL-asymmetry stored
in the quintessence field.
The final yield of the \BmL-asymmetry 
\begin{equation}\label{yield}
n_{\rm SM}/s = D\cdot \kappa \equiv D\cdot \frac{-A_\infty}{3.2\rho_0^{3/4}} \propto A_\infty
\end{equation}
(where $\rho_0\equiv 3\Hinf^2\Mpl^2$) can actually be calculated  either
numerically or, for a restricted parameter range, analytically via the integral in eq. (\ref{Ainfty})  using
an approximate WKB solution for $\chi(t)$~\cite{Bauer:2005vz} (see Fig. \ref{PhiAndChi} and Fig. \ref{Asym}),
\begin{equation}\label{kappaWKB}
 \kappa\approx-\frac{\mathcal{N}}{2}\sin(2\alpha_{0})\left(\frac{\chi_{0}}{H_{\rm {Inf}}}\right)^{\!2}\cdot  \left\{\rule[-12mm]{0mm}{24mm}\right.
	\begin{array}{l}
		 \!\!\!\!3.6\cdot 10^{-10}\frac{\phi_{0}}{10^{13}{\rm GeV}}\left(\frac{H_{\rm Inf}}{10^{12}{\rm GeV}}\right)^{\!\frac{1}{2}}
		\\[1mm] \hspace*{23mm} {\rm if}\ \phi_0^3 \gg \chi_0^2\phi_0, |V'(\phi_0)| \\
		\!\!\!\!1.7\cdot 10^{-8}\left(\frac{\xi_1}{7}\frac{V_{0}}{\rho_{0}}\right)^{\!\frac{1}{3}}\left(\frac{H_{\rm {Inf}}}{10^{12}{\rm GeV}}\right)^{\!\frac{7}{6}}
		\\[1mm] \hspace*{23mm} {\rm if}\ |V'(\phi_0)| \gg  \phi_0^3, \chi_0^3,
	\end{array}
\end{equation}
where $\mathcal{N}\equiv \mathcal{N}(\lambda_1,\lambda_2,g)$ contains the
the dependence on the coupling constants, with $\mathcal{N}\sim 1$ 
for $g^2/(8\pi)\sim\lambda_2/\lambda_1\ll\lambda_1\sim 1$~\cite{Bauer:2005vz}.
The analytic estimate agrees well with the numerical results (see Fig. \ref{Asym}) inside the respective domains of validity.
In the notation of eq. (\ref{yield}) $\kappa\propto A_\infty$ is the contribution
which depends on the dynamics of the quintessence and the mediating field,
and $D$ is a factor of proportionality which depends
on the expansion history of the universe
after inflation 
and can vary
in the range $1\gtrsim D \gtrsim 10^{-6}$ for various models of inflation and re/preheating~\cite{Bauer:2005vz}.
Thus, arriving at the observed value\footnote{Note that the \BmL-asymmetry and the
baryon asymmetry differ by an additional sphaleron factor of order one, see Ref.~\cite{Harvey:1990qw}.} $n_{\rm SM}/s \sim 10^{-10}$
is possible if the asymmetry parameter $\kappa$  is roughly in the range 
$10^{-10}\lesssim\kappa\lesssim10^{-4}$,
which is indeed the case for a broad range of values for the initial energy density
and VEV of the quintessence field (see right part of Fig. \ref{Asym}).

\section{Final Remarks}

An important issue in the context of complex quintessence models is to study the stability
against the formation of inhomogeneities, which could otherwise lead to the formation
of so-called Q-balls~\cite{Coleman:1985ki} and destroy the dark energy properties. 
Once the comoving asymmetry is frozen  one can estimate from eq. (\ref{BmLdensity})  the phase
velocity $\dot\theta_\phi$ which is necessary to yield an asymmetry $n_\phi/s\sim10^{-10}$,
\begin{equation}
\frac{|\dot\theta_\phi|}{H}=\frac{|n_\phi|}{2H|\phi|^2}\sim  
10^{-10} \frac{2\pi^2}{45}g_{*S}(T) \frac{T^3}{2H|\phi|^2} \lesssim 10^{-8} \frac{(HM_{\rm Pl})^{3/2}}{2H|\phi|^2}\ll 10^{-8}, 
\end{equation}
where we assumed $g_{*S}(T)\sim 100$ and $|\phi|\gtrsim \Mpl$.
Thus the field is moving extremely slowly in the radial direction compared to the
expansion rate of the universe, which is exactly the opposite limit as was studied
for example in the spintessence models~\cite{Boyle:2001du}. Quantitatively,
one can show~\cite{Kusenko:1997si} that there exist no growing modes for linear perturbations in $|\phi|$ and $\theta_\phi$ 
for any wavenumber $k$ provided that
$\dot\theta_\phi^2 < \frac{3H+2\dot\varphi/\varphi}{3H+6\dot\varphi/\varphi}V''$ 
(with $\varphi\equiv|\phi|$, $V''\equiv d^2V/d\varphi^2$).
Since the mass $V''\sim H^2$ of the quintessence field tracks the Hubble scale~\cite{Steinhardt:1999nw} 
and since $\dot\varphi/\varphi>0$ this inequality
is safely fulfilled once the tracking attractor is joined, and thus there are no hints for instabilities in this regime.
For a more detailed analysis including also the early moments of evolution as well as additional
particle processes we refer
to Ref.~\cite{Bauer:2005vz}.

Finally, we want to mention that, 
since the underlying Lagrangian is \BmL-symmetric, 
it offers a possibility to combine Dirac-neutrinos
with baryogenesis aside from the Dirac-leptogenesis mechanism~\cite{Dick:1999je}. 
Note that the lepton-asymmetry in the SM is
of opposite sign compared to Dirac-leptogenesis.
Furthermore, there is no specific lower bound on the reheating temperature like in thermal
leptogenesis~\cite{Davidson:2002qv}. 

In conclusion, the coupled leptonic quintessence model reviewed here
can account for the observed baryon asymmetry
of the universe without introducing new \BmL-violating interactions
below the inflationary scale by storing a lepton asymmetry in\enlargethispage{2mm}
the\nopagebreak[4] dark energy sector.

%
\subsection*{Acknowledgments}
%

---  This work is based on a collaboration with F.Bauer and M.T.Eisele and was supported
by the {}``Sonderforschungsbereich 375 f\"{u}r Astroteilchenphysik
der Deutschen Forschungsgemeinschaft''. 

\bibliographystyle{iopart-num}
\section*{References}
\bibliography{references-BAQuint}

\end{document}